\documentclass[twocolumn,superscriptaddress,showpacs]{revtex4-1}

\usepackage{amsmath,amsthm,amssymb,bm,hyperref,graphicx,color}

\newcommand{\be}{\begin{equation}}
\newcommand{\ee}{\end{equation}}

\newcommand{\lam}{\lambda}

\newcommand{\inti}{\int_{-\infty}^{+\infty}}
\newcommand{\g}{\gamma}
\newcommand{\6}{\partial}
\newcommand{\R}{\textsf{R}}

\newcommand{\T}{\textsf{T}}

\begin{document}

\title{Efficient Thermodynamic Description of Multi-Component One-Dimensional Bose Gases}

\author{Andreas Kl\"umper}
\affiliation{Fachbereich C – Physik, Bergische Universit\"at  Wuppertal,
42097 Wuppertal, Germany}
\author{Ovidiu I. P\^{a}\c{t}u}
\affiliation{Fachbereich C – Physik, Bergische Universit\"at  Wuppertal,
42097 Wuppertal, Germany}
\affiliation{Institute for Space Sciences, Bucharest-M\u{a}gurele, R
077125, Romania}


\begin{abstract}
We present a new method of obtaining nonlinear integral equations
characterizing the thermodynamics of one-dimensional multi-component gases
interacting via a delta-function potential. In the case of the repulsive
two-component Bose  gas we obtain a  simple system of two non-linear integral
equations,  allowing for an efficient numerical implementation, in contrast with the infinite
number of coupled equations obtained by employing the Thermodynamic Bethe Ansatz.
Our technique makes use of the Quantum Transfer Matrix and the fact that, in a
certain continuum limit, multi-component gases can be obtained from appropriate
anisotropic spin chains.
\end{abstract}

\maketitle

%%%%%%%%%%%%%%%%%%%%%%%%%%%%%%%%%%%%%%%%%%%%%%%%%%%%%%%%%%%%%%%%%%%%%%%%%%%%%%%%%%%%%%%%%%
{\it Introduction.} Recent advances in magnetic and optical trapping of ultracold
quantum gases have opened new possibilities in investigating the physics
of strongly interacting particles in one-dimension \cite{BDZ}. Coupled with the fact that
the strength of the atomic interaction can be controlled using magnetic Fesbach
resonances or state-dependent dressed potentials \cite{WWD}, these new experiments provide
a fertile ground for testing the theoretical predictions obtained from the study
of various integrable models solvable by Bethe Ansatz. As a result, we have witnessed
the experimental realization   of the Lieb-Liniger model \cite{LL} using bosonic
$^{87}Rb$ atoms \cite{MSK,LOH,P,KWW,PRD} and, moreover, the measured thermodynamics
\cite{A} is described very well by the Yang-Yang thermodynamics, which was developed
in \cite{YY}, using, what we call nowadays, the Thermodynamic Bethe Ansatz (TBA) \cite{T}.
In the case of integrable multi-component 1D gases, TBA produces an infinite number of
integral equations \cite{T,GLYZ}, which makes the extraction of physical information
and comparison with prospective experimental data difficult.
In this Rapid Communication, we propose a solution to this problem using the Quantum Transfer Matrix \cite{MS,K1},
and the connection between multi-component gases and anisotropic spin-chains. Our method
has the advantage of providing a finite number of non-linear integral equations (NLIE),
which are extremely suited for numerical computations, in stark contrast with the TBA result.
%%%%%%%%%%%%%%%%%%%%%%%%%%%%%%%%%%%%%%%%%%%%%%%%%%%%%%%%%%%%%%%%%%%%%%%%%%%%%%%%%%%%%%%%

{\it Model.} We consider a one-dimensional system of $M$  bosons, of equal
mass $m=1/2$,  with $n$ internal degrees of freedom,
interacting via a delta-function potential. The many-body Hamiltonian is

\begin{align}\label{Hc}
\mathcal{H}=-\sum_{i=1}^M\frac{\partial^2}{\partial x_i^2}+
2c\sum_{ i< j}\delta(x_i-x_j)-\sum_{i=1}^{n}\mu_i M_i\, ,
\end{align}
where $c$ is the coupling constant and we consider $\hbar=k_B=1$.
The first and the second term
in the Hamiltonian (\ref{Hc}), represent the kinetic and and the interaction
energy, while the third  is the Zeeman term, where $M_i$ is the number of
particles in the hyperfine state $|i\rangle$ and $\mu_i$ the respective
chemical potential. The interaction is attractive for
$c<0$ and repulsive for $c>0.$
The model is integrable and was solved in \cite{LL,Y1,G1,B1}.

Despite the fact that the model is integrable, computing the thermodynamics
is still an incredibly difficult task. The first attempt was done by Yang and
Yang \cite{YY} for the spinless bosons  using  the Thermodynamic Bethe Ansatz.
TBA has proved to be an extremely useful technique
and was used to obtain information about the thermodynamics  of various exactly
solvable models \cite{T}, however, it has one big shortcoming: for a large class
of models it produces an infinite number of coupled NLIE.  This is also the case for
our model, with the exception of the spinless case ($n=1$). From a very
simplified point of view this is due to the fact that the spectrum of the theory
contains infinitely many branches. The numerical implementation of this system
of equations requires various truncations,  which introduce an uncontrollable
source of numerical errors and makes the  extraction of relevant physical information
extremely difficult. Therefore, it is highly desirable to devise a numerically
efficient method, which provides a finite number of NLIE.

Fortunately, in the case of lattice models such a method exists.
Developed in \cite{MS,K1}, the Quantum Transfer Matrix (QTM) approach
was successfully applied to a large class of integrable spin-chains and
even models of strongly correlated electrons.
Within this approach, the thermodynamic properties of the model are
obtained from the  largest eigenvalue of the QTM,  which in the thermodynamic
limit gives the free energy.
Even though, there is no equivalent of the  QTM for continuum models,
it is well known \cite{KS,IK},  that a large class of integrable models
can be obtained from  lattice ones in suitable continuum limits. We say
that an integrable model is the continuum limit of a lattice
integrable model if, by performing  this limit in the Bethe equations and
energy spectrum of the lattice model one obtains the Bethe equations and
energy spectrum of the continuum model. The natural consequence is that the
thermodynamics of the continuum model can be obtained from the thermodynamics
of the lattice model, if we take the same limit. This method will be used
in this Rapid Communication, allowing  us to obtain an efficient thermodynamic description
of the two-component Bose gas (for the single component case see \cite{SBGA}).

In this publication we study and give results for the Hamiltonian (\ref{Hc})
in dimensionless units, however physical units can be restored easily. For
particles with mass $m$ and contact interaction strength $g$, see e.g. (1) in
\cite{GBT},
%\textcolor{green}{the units of temperature, chemical potential, effective
%  magnetic field (defined below),
the units of temperature, chemical potential,
magnetic field,
particle density and susceptibility, compressibility, heat capacity, and entropy per length are
$T_0=\hbar^2/(2 m a^2 k_B)$, $\mu_0=h_0= \hbar^2/(2 m a^2),$ $d_0=1/a$, $\chi_0=\kappa_0=2 m a/\hbar^2$,
$c_0=S_0=k_B/a$. The quantity $a$ is a length scale that can be chosen freely
yielding the dimensionless coupling constant $c=m g a/\hbar^2$ appearing in
(1). In all figures presented in this paper, physical data are shown in the
given units and for dimensionless coupling $c=1$ which is realized for any
parameter values of $m$ and $g$ with a suitably chosen $a=\hbar^2/(m g)$.

%%%%%%%%%%%%%%%%%%%%%%%%%%%%%%%%%%%%%%%%%%%%%%%%%%%%%%%%%%%%%%%%%%%%%%%%%%%%%%%%%%%%%%%%%%%%

{\it Method.} The general strategy is the following: first, we identify the spin-chain
from which we will obtain in the appropriate continuum limit our model of interest. The next step
is the investigation of the thermodynamics of the lattice model using the QTM technique,
which will provide a finite number of NLIE. Finally, the desired result is obtained by taking
the continuum limit in the equations for the lattice model. Let us make these general considerations more precise.
In the case of the  repulsive two-component  Bose gas, the relevant lattice model is the
$U_q(\widehat{sl}(3))$ Perk-Schultz spin-chain \cite{PS,BVV,dV,dVL,L}, with periodic boundary conditions
characterized by the Hamiltonian
\begin{align}\label{Hs}
H&=J\sum_{j=1}^L(
\cos\g\sum_{a=1}^3 e_{aa}^{(j)}e_{aa}^{(j+1)}+
\sum_{\substack{a,b=1\\ a \ne b}}^3
e_{ab}^{(j)}e_{ba}^{(j+1)}\\
&
-i\sin\g\sum_{\substack{a,b=1\\ a\ne b}}^3
\mbox{sign}(a-b)e_{aa}^{(j)}e_{bb}^{(j+1)})-
\sum_{j=1}^L\sum_{a=1}^3 h_a e_{aa}^{(j)}\, ,\nonumber
\end{align}
where $\g\in(0,\pi)$ determines the anisotropy, ($q=e^{i\g}$), $L$ is the number of lattice sites,
$J>0$ fixes the energy scale, $h_a$ are chemical potentials and
$e_{ab}^{(j)}= I^{\otimes j-1}\otimes e_{ab}\otimes I^{\otimes L-j}\, ,$
with $e_{ab}$ the $3$ by $3$ matrix with elements $(e_{ab})_{ij}
=\delta_{a i}\delta_{b j}$.  This Hamiltonian
is the sum of two terms, $H_b$ (in the brackets), and the chemical potential term
$H_c$, which  does not break the integrability of the model \cite{dV}.
The $U_q(\widehat{sl}(3))$ Perk-Schultz
spin-chain can be obtained as the fundamental spin model, (see \cite{KBI}), associated with the
trigonometric Perk-Schultz $\R$-matrix \cite{dV}, $\R(v)=\sum_{a,b,c,d=1}^3
\R^{ac}_{bd}(v)e_{ab}\otimes e_{cd},$
with
$
\R_{aa}^{aa}(v)=
\frac{\sin(\gamma+ v)}{\sin \gamma}\, ,
\R_{ab}^{ab}(v)  =
\frac{\sin v}{\sin\gamma}\, ,\  (a\ne b)\, ,
\R_{ab}^{ba}(v)  =
e^{i\, \mbox{sign}(a-b) v}\, ,\  (a\ne b)\, ,
$
and all the other elements zero.  The transfer matrix $\T(v)$ is the $3^L$ by $3^L$
matrix with elements
$
\T_{\mathbf b}^{\mathbf a}(v)=\sum_{\{{\mathbf c} \}}\prod_{j=1}^L
\R_{c_j\,\ \  b_j}^{c_{j+1}a_j}(v)\, ,
$
where $\mathbf a,\mathbf b$, and $\mathbf c$
 are multiple indices, {\it i.e., }$ \mathbf a=(a_1,\cdots,a_L)$
and the sum is over all $\mathbf c$ with $c_1=c_{L+1}$.
The $H_b$ part of the Hamiltonian (\ref{Hs}) can be obtained as the logarithmic derivative
of the transfer matrix $H_b=J\sin\gamma\frac{\partial \ln \T(v)}{\partial v}|_{v=0}.$
In order to treat the model at finite temperature we introduce the $3^N$ by $3^N$
Quantum Transfer Matrix with elements
\begin{align*}
&(\T_{QTM})_{\mathbf b}^{\mathbf a}(v)=\sum_{\{\mathbf{c}\}}e^{\beta h_{c_1}}
\prod_{j=1}^{N/2}\R_{c_{2j}\ \ \  b_{2j}}^{c_{2j+1}a_{2j}}(v+iu)\\
&\ \ \ \ \ \ \ \ \ \ \ \ \ \ \ \ \ \ \ \ \ \ \ \ \ \ \ \ \ \ \ \ \ \ \ \ \ \
\times \tilde\R_{c_{2j-1}b_{2j-1}}^{c_{2j}\ \ a_{2j-1}}(iu-v)\, ,
\end{align*}
where $c_1=c_{N+1},$ $\tilde\R_{bd}^{ac}(v)=\R_{da}^{cb}(v),$
$u=-J\beta\sin\gamma/N,$ with $N$ the Trotter number and $\beta=1/T$.
The largest eigenvalue of the QTM, from which the free energy of the model can
be obtained, has the form:  $\Lambda_{QTM}(v)=\sum_{j=1}^3\lambda_j(v)$ \cite{KWZ},
with
\[
\lam_j(v)=\phi_-(v)\phi_+(v)\frac{q_{j-1}(v-i\g)}{q_{j-1}(v)}\frac{q_{j}(v+i\g)}{q_{j}(v)}e^{\beta h_j}\, ,
\]
where $ \phi_{\pm}(v)=\left(\frac{\sinh(v\pm iu)}{\sin \g}\right)^{N/2}\,  $ and
\[
q_j(v)=\left\{\begin{array}{lr}
               \phi_-(v) & j=0\, \\
               \prod_{r=1}^{N/2}\sinh(v-v_{r}^{(j)})& j=1,2\, \\
%%
%%               \phi_+(v) & j=n\, \\
%%
%%
               \phi_+(v) & j=3\, .\\
              \end{array}\right.
\]
The two sets of parameters, $\{v_r^{(j)}\},$ are the Bethe roots of the QTM and satisfy the Bethe
equations $\lam_j(v^{(j)}_r)/\lam_{j+1}(v^{(j)}_r)=-1\, ,\  r=1,\cdots,N/2\, . $
The Bethe roots  are distributed  in certain strips of the complex plane which
are independent of temperature and Trotter number. This type of specific distribution
allows for the definition of auxiliary functions and the use of Cauchy's theorem in
deriving NLIE for the largest eigenvalue \cite{KB,K1,DDV,K2,K3}. Our results, which
in the continuum limit produce Eqs.~\ref{aux} and \ref{p}, can be understood as the
multi-component generalization of  the equations presented in \cite{K3}.

%%%%%%%%%%%%%%%%%%%%%%%%%%%%%%%%%%%%%%%%%%%%%%%%%%%%%%%%%%%%%%%%%%%%%%%%%%%%%%%%%%%%%%%%

{\it Main result.} In order to obtain the thermodynamics of  the  two-component
Bose gas we have to perform a particular continuum limit in the NLIE obtained for
the   $U_q(\widehat{sl}(3))$ spin-chain.
The spin-chain is characterized by the following parameters: lattice constant $\delta$,
number of lattice sites $L$, anisotropy $\g$, strength of the interaction $J$,
and chemical potentials $h_1,h_2,h_3$. The two-component Bose gas is obtained
by performing the limit $\g=\pi-\epsilon\, , \delta\rightarrow O(\epsilon^2)\, ,
L\rightarrow O(1/\epsilon^2)\, , J\rightarrow O(1/\epsilon^4)\, ,h_1\rightarrow
O(1/\epsilon^2)$ with $\epsilon\rightarrow 0$. Performing this limit in the Bethe equations
and energy spectrum of the  $U_q(\widehat{sl}(3))$ spin-chain, we obtain the
Bethe equations and energy spectrum of a two-component Bose gas characterized by
the parameters: length $l=L\delta$, coupling constant $c=\epsilon^2/\delta$, mass
of the particles $m=J\delta^2$, and chemical potentials $\mu_1=J\epsilon^2+h_2-h_1
\, , \mu_2=h_3-h_2$.
In order to simplify the formulae, we are going to consider $J\delta^2=1$
(mass of the particles equal to $1/2$) and introduce $\mu=(\mu_1+\mu_2)/2$ and
an effective magnetic field $h=(\mu_1-\mu_2)/2$, where
$\mu_{1,2}$ are the chemical potentials of the spin up
and down particles. We are now ready to state the main result
of this Rapid Communication. The thermodynamics of the  two-component repulsive Bose gas is
completely characterized by the following system of nonlinear integral equations:
\begin{subequations}\label{aux}
\begin{align}
\ln a_1(k)&=-\beta(k^2-\mu-h)\nonumber\\
&\ \   +[K_0*\ln A_1](k)+[K_1*\ln A_2](k+i\epsilon)\, ,\label{a}\\
\ln a_2(k)&=-\beta(k^2-\mu+h)\nonumber\\
&\ \  +[K_2*\ln A_1](k-i\epsilon)+[K_0*\ln A_2](k)\, ,
\end{align}
\end{subequations}
where  $ A_i(k)=1+a_i(k)\, ,$ $K_0(k)=2c/(k^2+c^2)\, ,$
 $K_1(k)=c/[k(k+ic)]\, ,K_2(k)=c/[k(k-ic)]$ and $[f*g](k)=\frac{1}{2\pi}\inti
f(k-k')g(k')\, dk'.$ The grandcanonical potential per length
is given by
\be\label{p}
\phi=-\frac{T}{2\pi}\inti \left(\ln A_1(k)+\ln A_2(k)\right)\, dk\, ,
\ee
from which all the other thermodynamic quantities can be obtained.
Eqs.~\ref{aux} and \ref{p} are new in the literature, and to our knowledge this
is the first example of an efficient thermodynamic description for a multi-component
continuum integrable model at all values of the relevant parameters (temperature,
chemical potentials and coupling constant).

\begin{figure}[h]
\includegraphics[width=\linewidth, height=63mm]{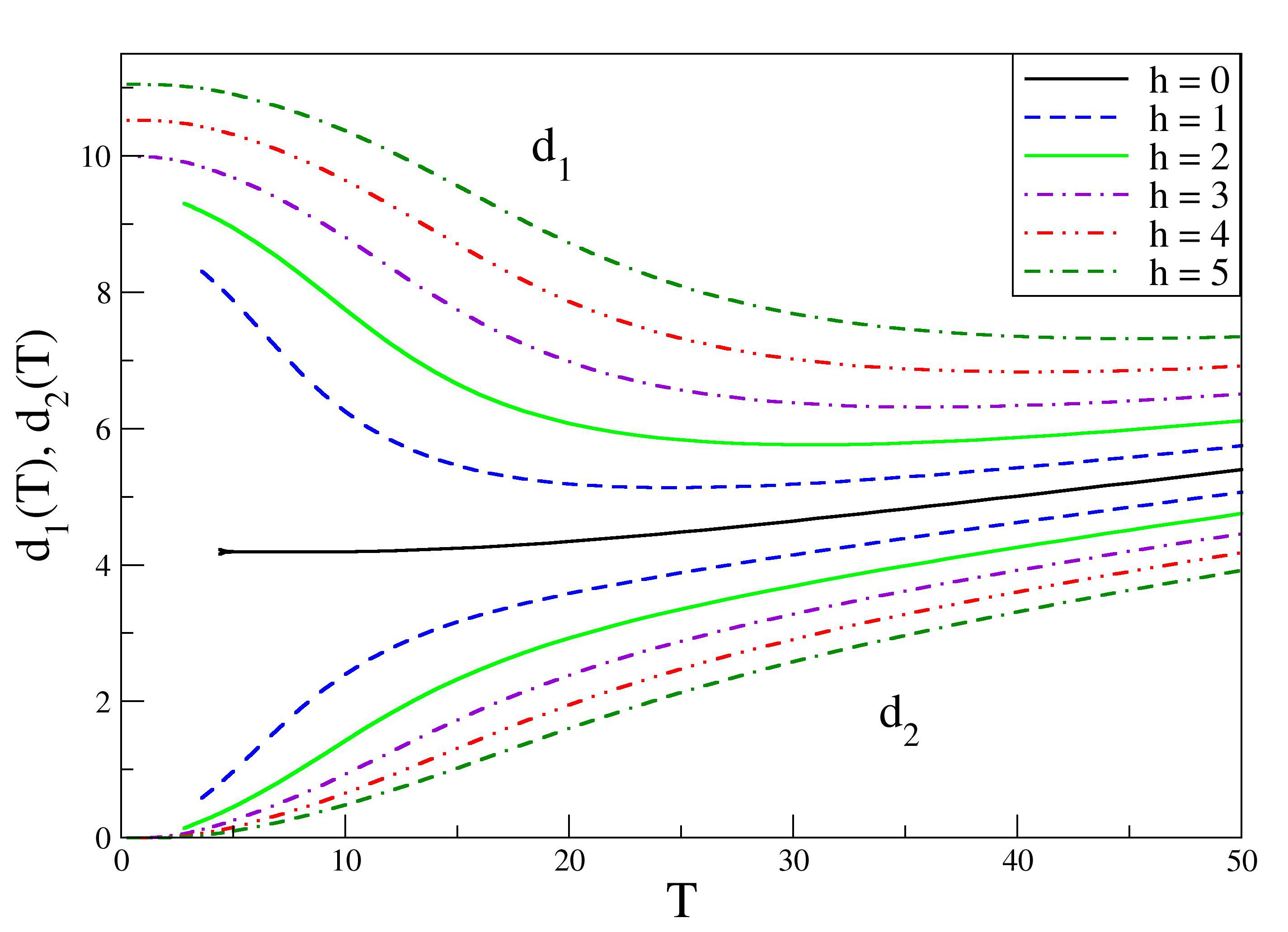}
\caption{(Color online) Population densities $d_1$, $d_2$
in the upper, lower part of the figure as functions of temperature $T$
for $\mu=15,\, c=1$ and various effective magnetic fields $h$
(in units of $d_0$, $T_0$, $\mu_0$ and $h_0$).  }
\label{populations}
\end{figure}
\begin{figure}[h]
\includegraphics[width=\linewidth,height=63mm]{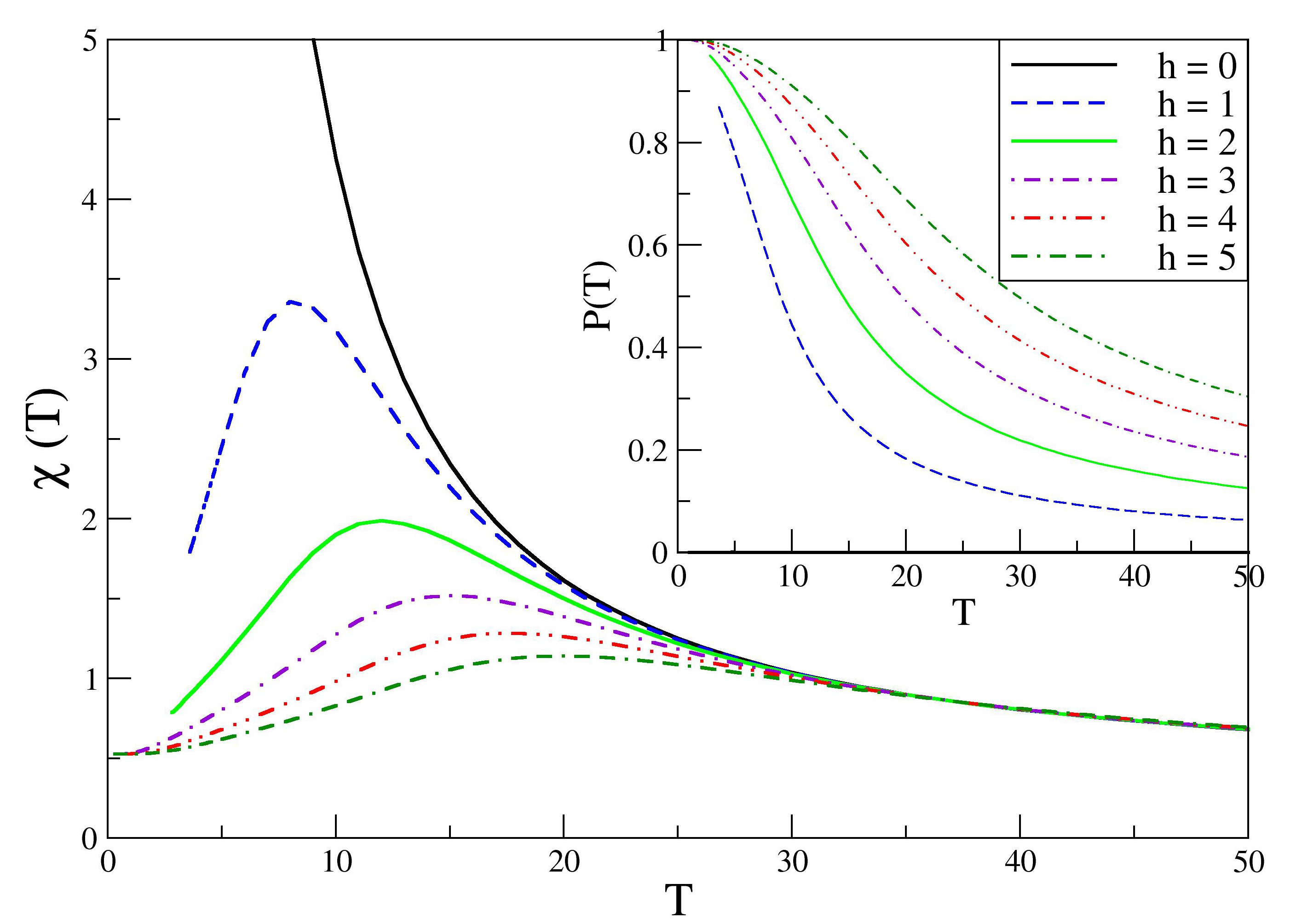}
\caption{(Color online)
Susceptibility $\chi$ as a function of temperature $T$
for $\mu=15, $ $c=1$ and various effective magnetic fields $h$
(in units of $\chi_0$, $T_0$, $\mu_0$ and $h_0$).
Inset: polarization as a function of temperature  for the same parameters. In the absence of the magnetic field
the polarization is zero.}
\label{susceptibility}
\end{figure}

{\it Analytic limits.} First, we will consider the limit $c\rightarrow 0.$
Then, we have $ \lim_{c\rightarrow 0}c/(k^2+c^2)=\pi\delta(k)\, $ and
$ \lim_{c\rightarrow 0}c/[k(k \pm ic)]=0\, $
and the integral equations decouple
\begin{align*}
\ln a_1(k)&=-\beta(k^2-\mu-h)+\ln(1+a_1(k))\, ,\\
\ln a_2(k)&=-\beta(k^2-\mu+h)+\ln(1+a_2(k))\, .
\end{align*}
These equations are easily solvable, and we find
\[
\phi=\frac{T}{2\pi}\inti \ln\left[(1-e^{-\beta(k^2-\mu-h)})(1-e^{-\beta(k^2-\mu+h)})\right] dk\, ,
\]
which is exactly the grandcanonical potential of two
noninteracting bosonic gases at different chemical potentials.

In the case of an extremely strong magnetic field, $h\rightarrow \infty$, we expect
that the system will become fully polarized and we should obtain the Yang-Yang
thermodynamics \cite{YY} of the repulsive one-component
Bose gas. In this limit
$a_2(k)\sim 0$ and we obtain
\[
\ln a_1(k)=-\beta(k^2-\mu-h)+[K_0*\ln(1+a_1)](k)\, ,
\]
and $ \phi=-\frac{T}{2\pi}\inti \ln(1+a_1(k))dk\, ,$
which is  the result obtained in \cite{YY}.
The same result is obtained in the low temperature limit  ($T\ll\mu, h, c$), when
the magnetic field is finite and  fixed which shows the ferromagnetic (fully polarized)
nature of the groundstate.

\begin{figure}
\includegraphics[width=\linewidth, height=63mm]{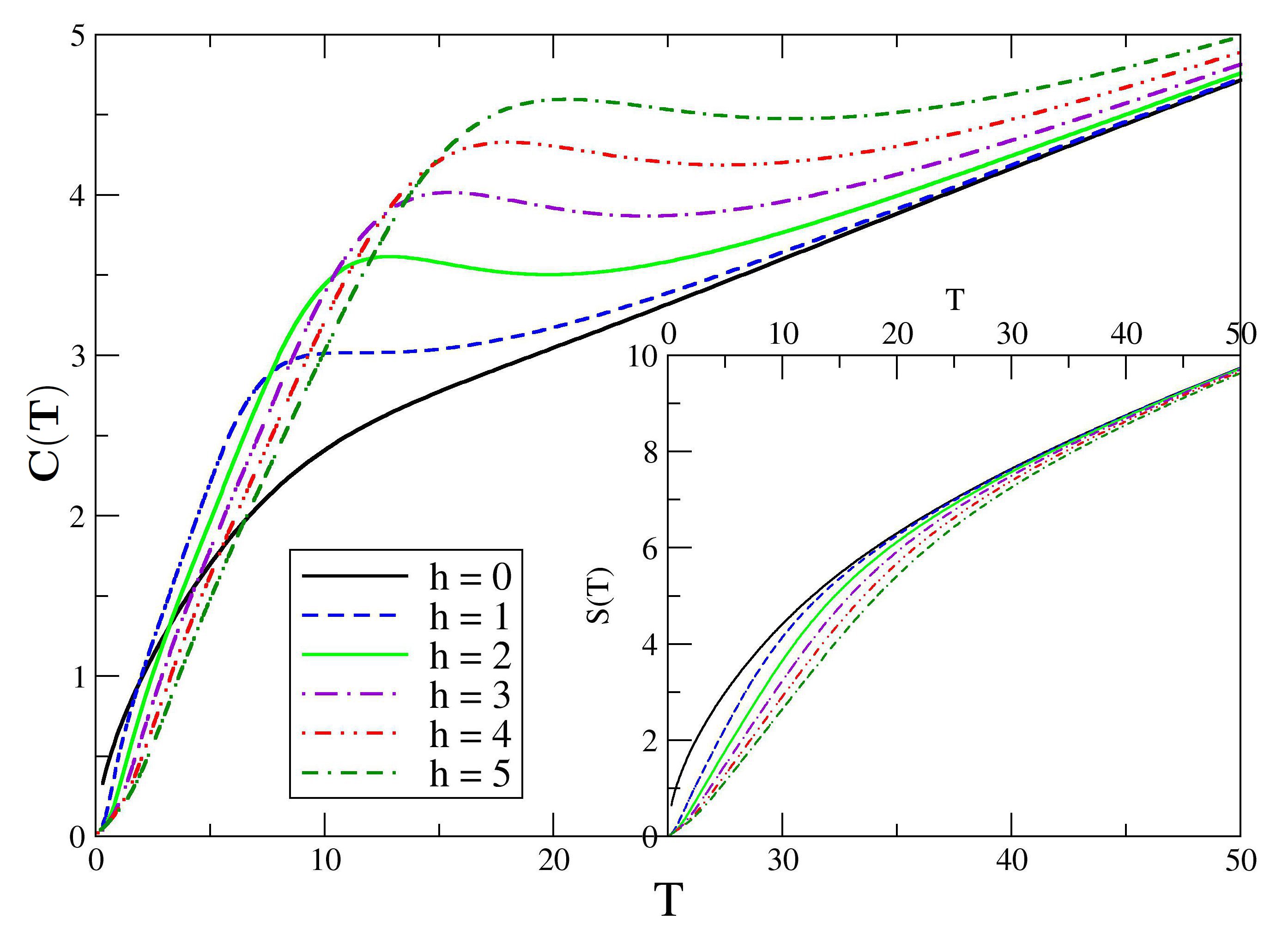}
\caption{(Color online)
Specific heat $C$ as a function of temperature $T$
for $\mu=15,\, c=1$ and various effective magnetic fields $h$.
Inset: entropy $S=-\frac{\6 \phi}{\6 T}$ as a function of
temperature for the same parameters. (All quantities in units of
$c_0$, $T_0$, $\mu_0$, $h_0$ and $S_0$.)}
\label{heat}
\end{figure}

\begin{figure}[t]
\includegraphics[width=\linewidth, height=63mm]{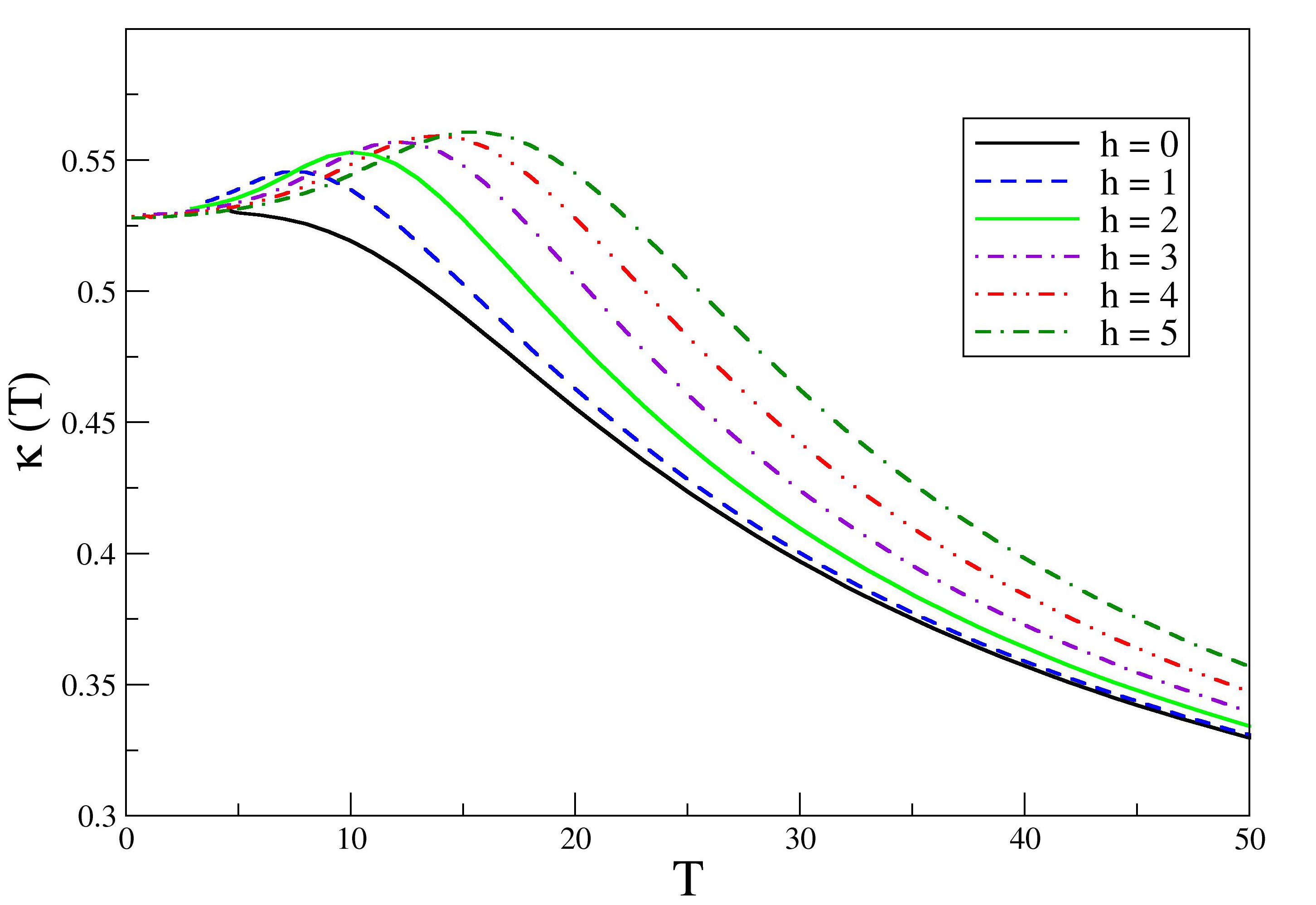}
\caption{(Color online)
Compressibility $\kappa$ as a function of temperature $T$
for $\mu=15,\, c=1$ and various effective magnetic fields $h$
(in units of $\kappa_0$, $T_0$, $\mu_0$ and $h_0$).}
\label{compressibility}
\end{figure}

As we have mentioned, the numerical implementation of the infinite number
of coupled NLIE obtained with the help of TBA \cite{GLYZ} is extremely difficult
and encounters serious problems in the regime of low magnetic field
$(h\ll T, \mu,c),$ or  low temperature limit $(0<T\ll h,\mu,c).$
In contrast, Eqs.~\ref{aux} are easily implementable and provide
reliable results in a larger domain  of the parameters space. We have
checked our results with available numerical data obtained by using the
much more involved TBA equations \cite{CKB} and found perfect agreement.
In our case, the regime in which the accuracy is decreasing, is given by the
low temperature and low magnetic field limit $(T\rightarrow 0, h\rightarrow 0).$
This is a consequence of
%the \textcolor{red}{ phase transition to the fully polarized,
%ferromagnetic groundstate, }which takes place
%at zero temperature when a magnetic field is switched \textcolor{red}{ on}.
a first order phase transition resp.~coexistence of fully
  polarized phases at $T$ and $h$ equal to 0.
It can be seen clearly in Figs.~\ref{populations} and \ref{susceptibility} that,
in the absence of the magnetic field the population levels,
$d_i=-\left(\frac{\6 \phi}{\6 \mu}+(-1)^{i-1}
\frac{\6 \phi}{\6 h}\right)/2\, ,\  i=1,2$, are equal and, consequently,
the polarization defined as $P=(d_1-d_2)/d$ with $d=d_1+d_2$ is zero. In the presence of
the magnetic field the groundstate is ferromagnetic, therefore, the
polarization at $T=0$ is $+1$, for any positive field ($-1$ for negative
field) and it decreases at higher temperatures. The ferromagnetic nature of the
groundstate, which can also be seen   in the $T^{1/2}$ behavior of the specific heat
$C= T\frac{\6 S}{\6 T}$
at zero magnetic field
(see Fig.~\ref{heat}), is in accordance with a more general theorem of Eisenberg
and Lieb  \cite{EL} on systems with  spin-independent interactions (see also \cite{GBT}).
Another consequence of the phase transition is the divergence of the zero-field
susceptibility, $\chi=-\frac{\6^2\phi}{\6 h^2}$, a feature which
can be seen in Fig.~\ref{susceptibility}. The specific heat,
magnetic susceptibility, and compresibility, $\kappa=-\frac{\6^2\phi}{\6\mu^2}$,
(see Fig. \ref{compressibility}), present a nonmonotonic behavior with local
maxima shifting to higher temperatures as the magnetic field increases.

In conclusion, we have presented a new method of obtaining a finite number of  NLIE
characterizing  the thermodynamics of integrable multi-component  1D Bose gases, which
allows for an efficient numerical treatment and  has significant advantages over
the TBA result.
Fermionic gases can be treated in a similar fashion, but in this case
the relevant lattice model for the $n$-component system is the $U_q(\widehat{sl}(n|1))$
Perk-Schultz spin-chain. The derivation and a  detailed analysis of
Eqs.~\ref{aux} and \ref{p} will be presented in a future publication.

{\it Acknowledgments.}  The authors would like to thank D.V. Averin, H. Boos, J.S. Caux and
F. G\"ohmann for valuable discussions. The authors are particularly grateful to
J.S. Caux for the use of his program for a comparison of the numerical data
obtained by the fundamentally different TBA and QTM methods.
O.I.P. acknowledges the financial support from the VolkswagenStiftung
and the LAPLAS3-PN 03 39 program of the
Romanian National Authority for Scientific Research.

\end{document}